\documentclass[final,5p,times,twocolumn]{elsarticle}

\usepackage{graphicx}

\usepackage{amssymb}
\usepackage{amsmath}

\journal{Nuclear Instruments and Methods A}

\begin{document}

\begin{frontmatter}

\title{New Approaches for Improvement of TOF-PET}

\author[A]{S. E. Brunner}
\author[A]{L. Gruber}
\author[A]{J. Marton}
\author[A]{K. Suzuki}
\author[B]{A. Hirtl}
\address[A]{Stefan Meyer Institute for Subatomic Physics, Austrian Academy of Sciences, Vienna, Austria}
\address[B]{Department of Radiology and Nuclear Medicine, Medical University of Vienna, Vienna, Austria}

\begin{abstract}
We present results of simulations on the influence of photon propagation and the Cherenkov effect on the time resolution of LSO:Ce scintillators. The influence of the scintillator length on the coincidence time resolution is shown. Furthermore, the impact of the depth of interaction on the time resolution, the light output and the arrival time distribution at the photon detector is simulated and it is shown how these information can be used for time walk correction.
\end{abstract}

\begin{keyword}
positron emission tomography (PET)
\sep
time-of-flight (TOF)
\sep
time resolution
\sep
Cherenkov effect
\sep
depth of interaction (DOI)
\sep
time walk
\end{keyword}
\end{frontmatter}
\section{Introduction}
Advances in detector technology led to the construction of time-of-flight (TOF) positron emission tomography (PET) scanners, resulting in enhanced signal-to-noise ratio. Goal of this work is the improvement of time resolution of scintillation detectors for PET, by a better understanding of photon propagation inside scintillators and the influence of the Cherenkov effect.

The time resolution of scintillation detectors depends on several factors. It can be distinguished between the time resolution of the scintillator, the photon detector and the readout electronics. The magnitude of time resolution of the scintillator itself has two major origins: statistical processes of the scintillation ($\sigma_{\textrm{\footnotesize{statistics}}}$) and the photon propagation from the point of interaction to the photon detector ($\sigma_{\textrm{\footnotesize{propagation}}}$). The statistical processes are in principle influenced by scintillation rise- and decay-times and the light yield. The photon propagation process is influenced by factors, such as refractive index, scintillator geometry and surface finishing.

Cherenkov photons in scintillators are emitted by electrons, ionised by incident 511\,keV photons and propagate faster than the speed of light $c/n$ in the scintillator, with \textit{c} and \textit{n} being the speed of light in vacuum and the refractive index, respectively. Making use of the Cherenkov effect is very promising for TOF-PET detectors, as the time spread of this process is smaller than for scintillation in inorganic materials \cite{Dolenec2010,Lecoq2010}. The direction of the Cherenkov photons can be described as a cone relative to the electron motion. The opening angle is determined by electron velocity and the refractive index of the crystal. Since the electron can be scattered to any direction, the direction of the photons is quasi-random. The number of emitted Cherenkov photons for LSO:Ce is about 15 per 511\,keV photon \cite{Lecoq2010}.

Monte Carlo simulations were performed to investigate the impact of photon propagation and the Cherenkov effect inside scintillators on the time resolution. These simulations will be described in the following sections.

\section{Simulation Setup}
For the simulations Geant\,4 \cite{Geant4} was used. LSO:Ce was chosen as scintillator material, since this is a common scintillating material for PET. The optical properties, such as, refractive index and transmission spectrum were taken from \cite{Mao2007}. For the light yield, decay and rise time, typical values of $3\cdot10^4$ photons/MeV, 40\,ns and 100\,ps were chosen, respectively. In order to evaluate only the effect of photon production and photon propagation, the detection efficiency of the photon detectors was set to 100\,\% over all wavelengths. Instead of simulating a positron source, photons with energies of 511\,keV were generated in a point source and emitted into a defined direction. The photon detectors recorded arrival time, wavelength and creation process of the photons. Compton scattered events were discriminated.

\section{Coincidence Time Resolution}
\label{sec:CTR}
Besides photon statistics, the coincidence time resolution (CTR) of PET-like detector systems is influenced by variations of the depth of interaction (DOI) in the opposing scintillators, resulting in differences of the photon propagation lengths to the photon detectors. This effect can be reduced by shortening the crystals, however, leading to decreasing detection efficiency for 511\,keV photons.

\begin{figure}[b] 
\centering 
\includegraphics[width=1.\columnwidth,keepaspectratio]{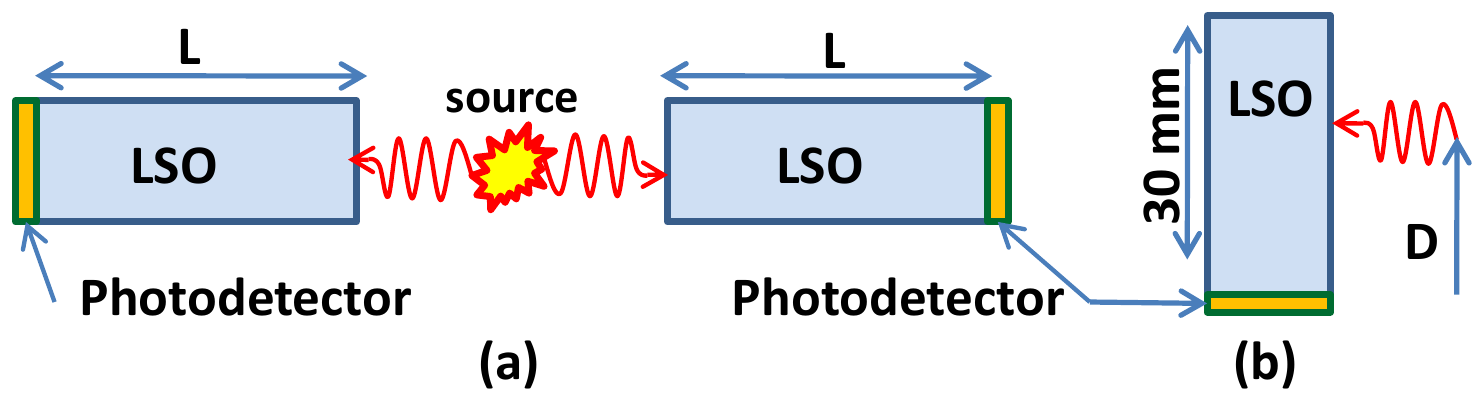}
\caption{Setup for the Geant\,4 simulations. The setup (a) was used for the simulations on the coincidence time resolution, where the length of the LSO crystal \textit{L} was varied from 1\,mm to 30\,mm. The source of the 511\,keV photons was placed in between the crystals. Configuration (b) shows the setup for the simulations on the influence of the DOI on the time spread of photons at the detector due to photon propagation and the photon output of a scintillator. In this case the crystal is 30\,mm long and \textit{D} is the distance of the 511\,keV photon source, which varies from 0.25 to 29.75\,mm.}
\label{fig:CoincidenceSetup}
\end{figure}

To evaluate the coincidence time resolution, a basic coincidence setup with two finger-like scintillators, each connected to one photon detector, was simulated, see figure \ref{fig:CoincidenceSetup} (a).  The crystal sizes were $3 \times 3 \times Z$\,mm$^3$ with Z ranging from 1\,mm to 30\,mm.

\begin{figure}[hbt] 
\centering 
\includegraphics[width=0.7\columnwidth,keepaspectratio]{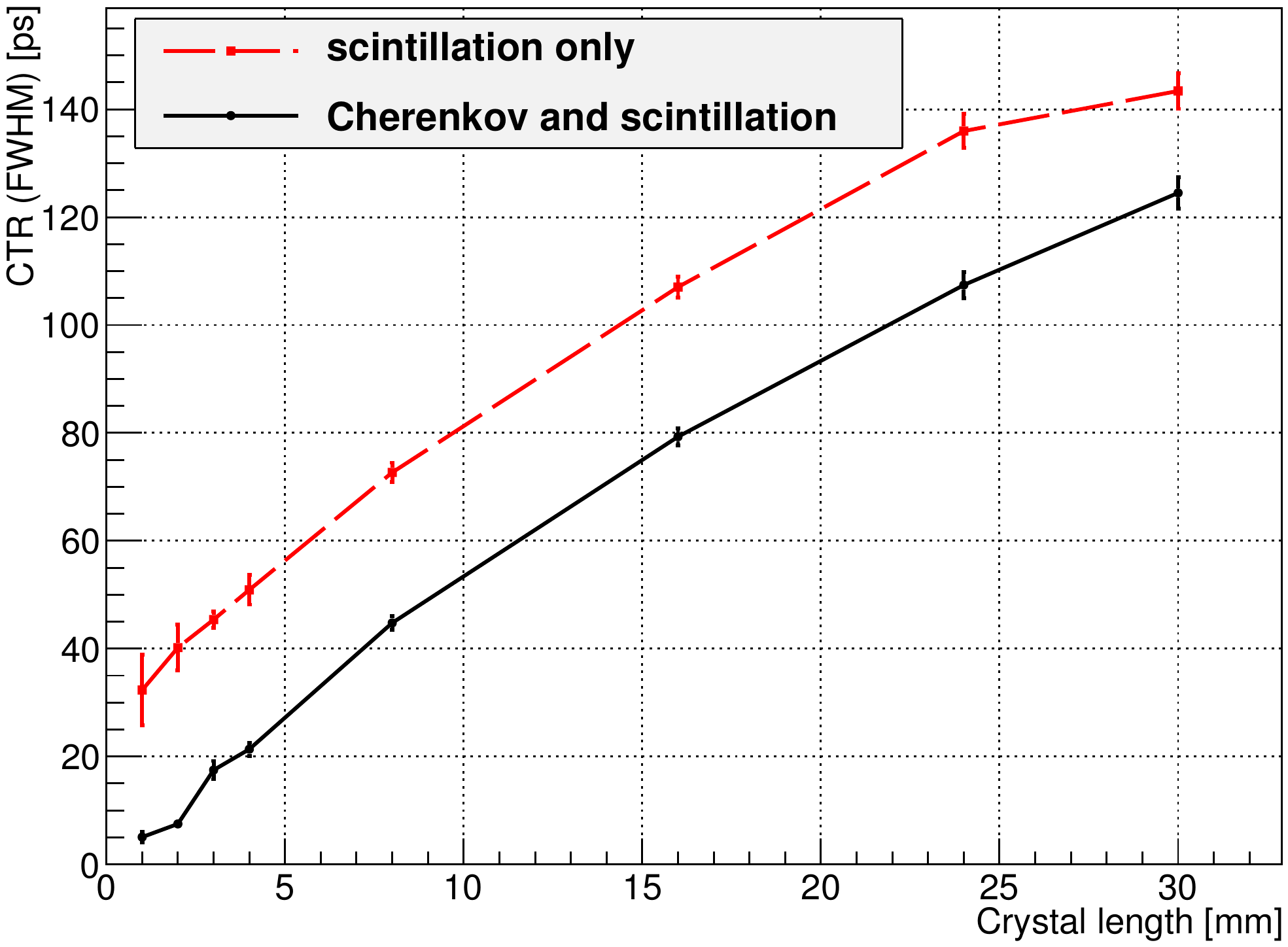}
\caption{CTR for various crystal lengths. The dashed curve indicates the CTR when only scintillation photons are detected. The solid line shows the CTR if Cherenkov photons are included in the detection.}
\label{fig:CTR_LSO}
\end{figure}

In figure \ref{fig:CTR_LSO}, the simulated CTR is shown for various crystal lengths. Two curves are plotted, one representing the detection of scintillation photons only, the other one showing the improvement of time resolution if the detection of Cherenkov photons is included. For both lines the dependency on the crystal lengths is clearly visible. For crystal lengths from 1\,mm to 30\,mm the CTR ranges from 32\,ps to 144\,ps FWHM for scintillation and from 12\,ps to 125\,ps FWHM by including the Cherenkov effect, respectively. The reason for this behaviour is the decrease in localisation of the 511\,keV photon interaction inside both of the crystals, simply due to the increasing size of the crystals. This uncertainty of localisation causes increasing time spread as a function of the crystal length. The impact of the Cherenkov effect on the CTR is clearly visible in figure \ref{fig:CTR_LSO}.

Although time resolution improves with decreasing crystal lengths, longer crystals, providing reasonable sensitivity to the 511\,keV photons, are used for real PET systems. The simulated detection efficiency of coincidences ranges from 2\,\%, for $Z=1$\,mm to above 50\,\% for $Z=30$\,mm. For real TOF-PET scanners a trade-off between time resolution and sensitivity has to be made.

\section{DOI and Time Resolution}
\label{sec:photontransport}
\begin{figure*}[bt] 
\centering 
\includegraphics[width=0.33\textwidth,keepaspectratio]{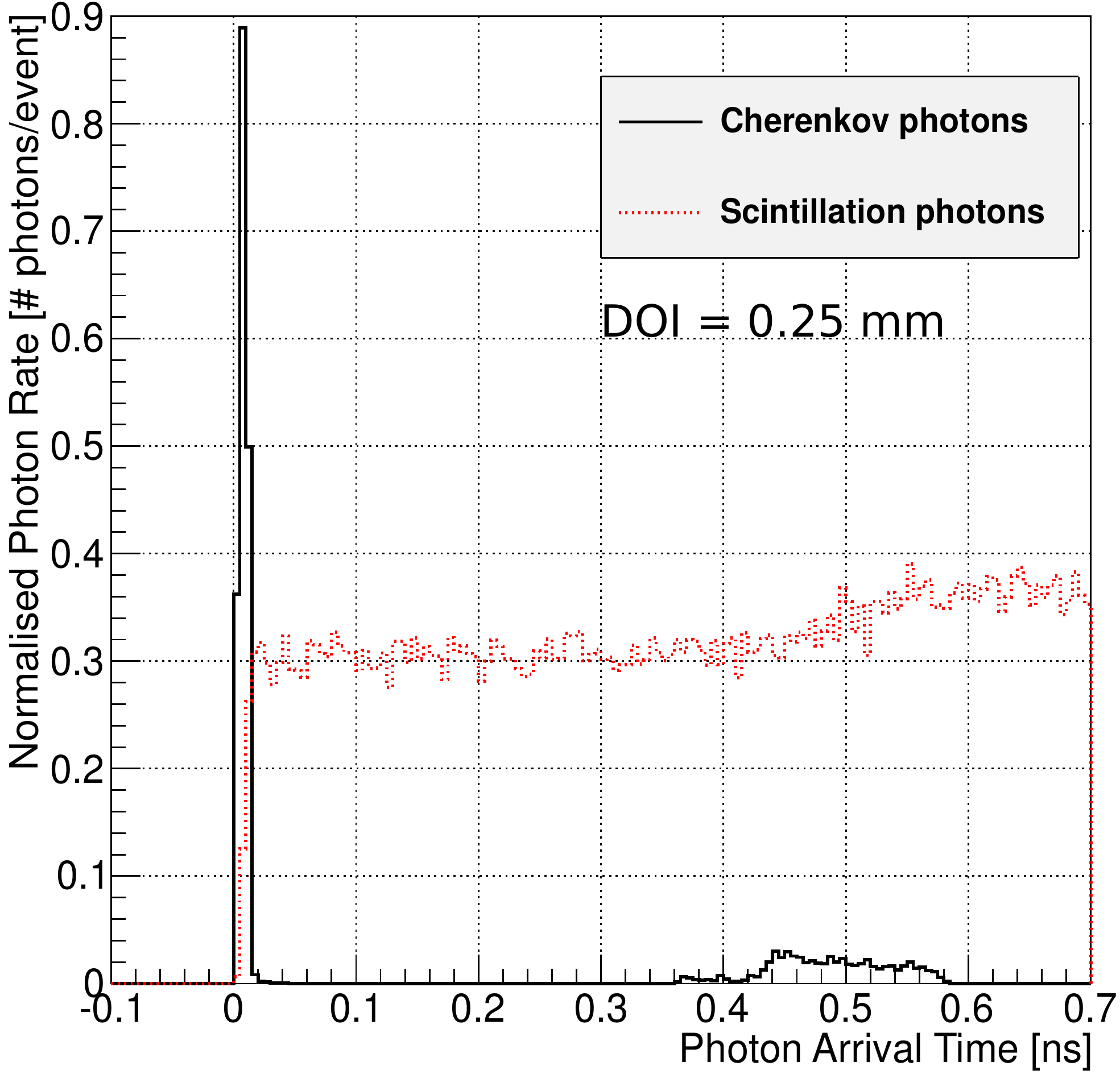}
\includegraphics[width=0.33\textwidth,keepaspectratio]{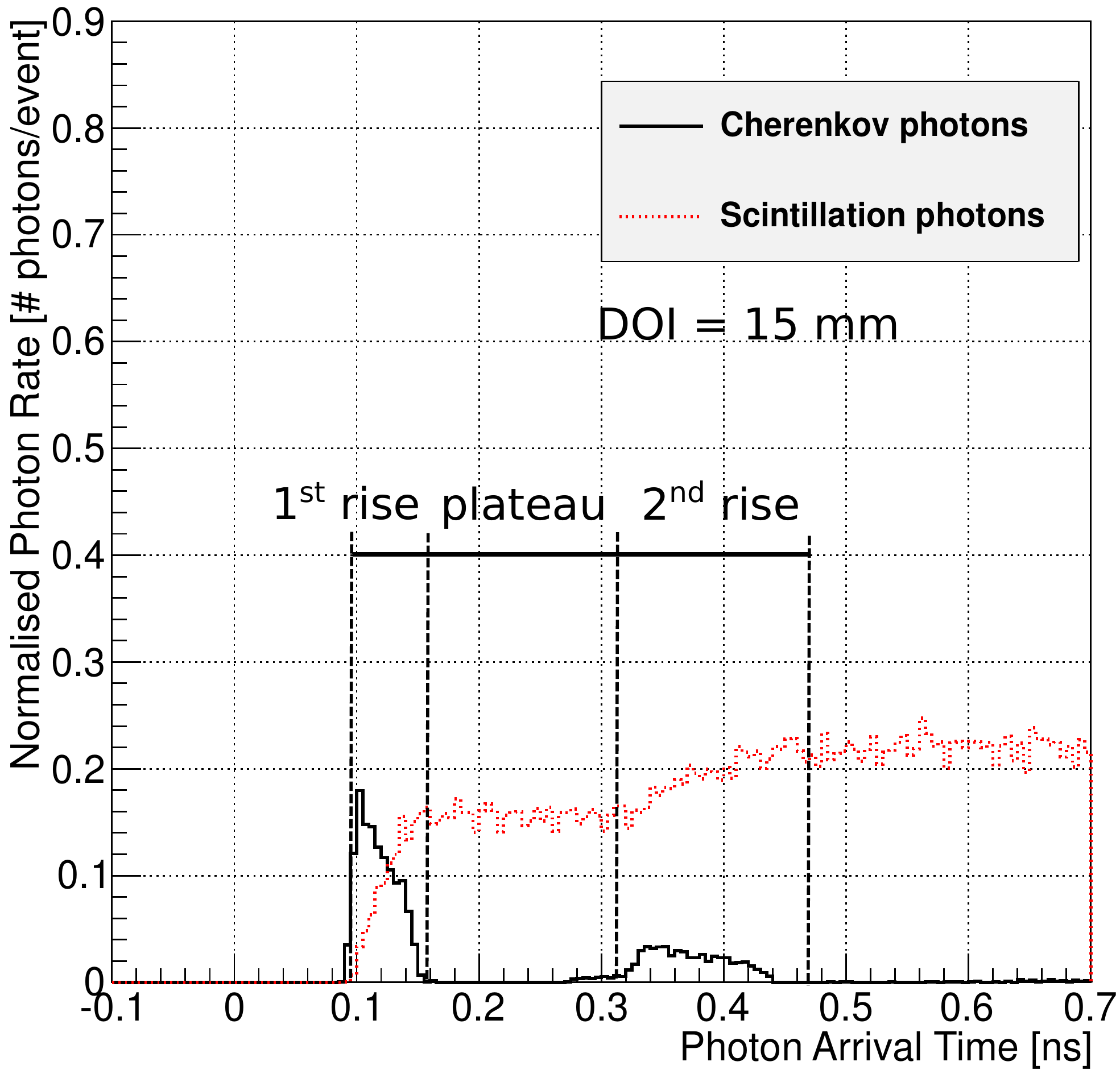}
\includegraphics[width=0.33\textwidth,keepaspectratio]{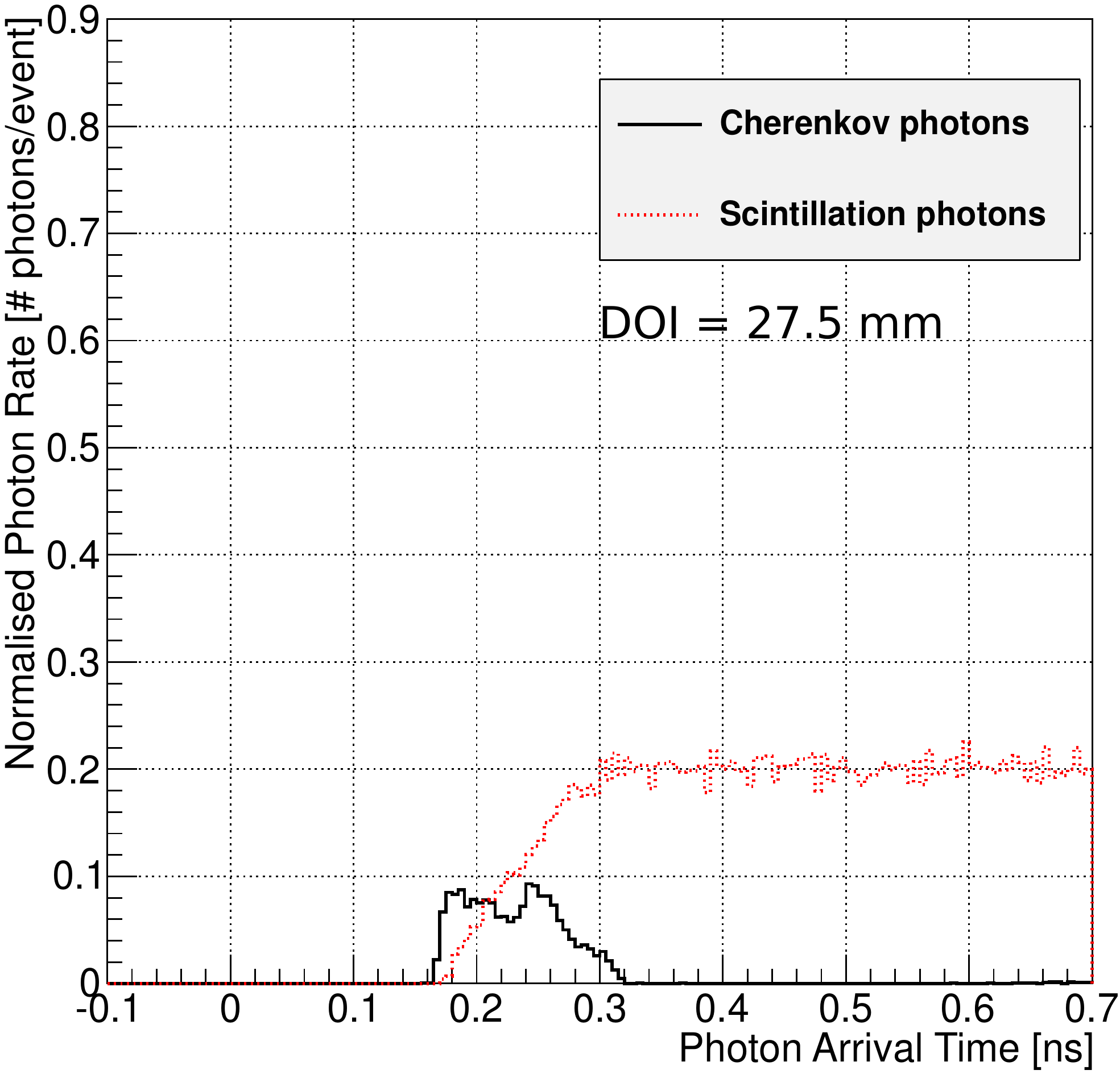}
\caption{Normalised photon arrival rates at the photon detector for various DOIs. The photon rates are discriminated by their creation process, which is either the scintillation or the Cherenkov effect. Normalisation was done using the number of detected 511\,keV photons within the photopeak in the energy spectrum (i.\,e. an event). As bin size 10\,ps was chosen. In the plot for a DOI of 15\,mm the $1^{\textrm{\footnotesize{st}}}$ rise, the intermediate plateau and the $2^{\textrm{\footnotesize{nd}}}$ rise are marked for the arrival rate of scintillation photons.}
\label{fig:ArrivalTimes}
\end{figure*}

In the following, the impact of the DOI on the photon arrival rates and the photon output of the scintillator at the photon detector will be discussed. For the simulations, the setup of figure \ref{fig:CoincidenceSetup} (b) was used. The size of the simulated crystal was $3 \times 3 \times 30$\,mm$^3$, and was connected to a photon detector. A source of 511\,keV photons was placed at the side of the scintillator and the distance \textit{D} of the source relative to the photon detector was varied over the whole crystal length from 0\,mm to 30\,mm. By knowing the distance \textit{D}, the DOI is determined.

The simulated photon arrival rates for three distances \textit{D} at the photon detector are shown in figure \ref{fig:ArrivalTimes}. For the photon arrival rates coming from the scintillation process, a fast rise can be seen at early times, followed by an intermediate plateau until a second, smaller and slower increase of the photon rate is visible. The width of the plateau is directly related to the DOI of the penetrating 511\,keV photons and vanishes for DOIs reaching the length of the scintillator.

The reason is originated in the isotropic emission of scintillation photons. The photons, emitted towards the photon detector form the first rise of the number of scintillation photons and the consecutive plateau. The second rise is caused by photons emitted away from the photon detector, getting reflected at the end of the scintillator and reaching the photon detector with a delay, depending on their travel path.

For the Cherenkov photons this effect is more obvious, since the duration of the Cherenkov process is shorter, compared to the  scintillation process. The Cherenkov photons form two subsequent peaks with a distance and width proportional to the effective travel path inside the scintillator. Compared to the scintillation photons, the rates of Cherenkov photons form sharp peaks, providing accurate time information of the interactions of the 511\,keV photons.

\begin{figure}[bt] 
\centering 
\includegraphics[width=0.85\columnwidth,keepaspectratio]{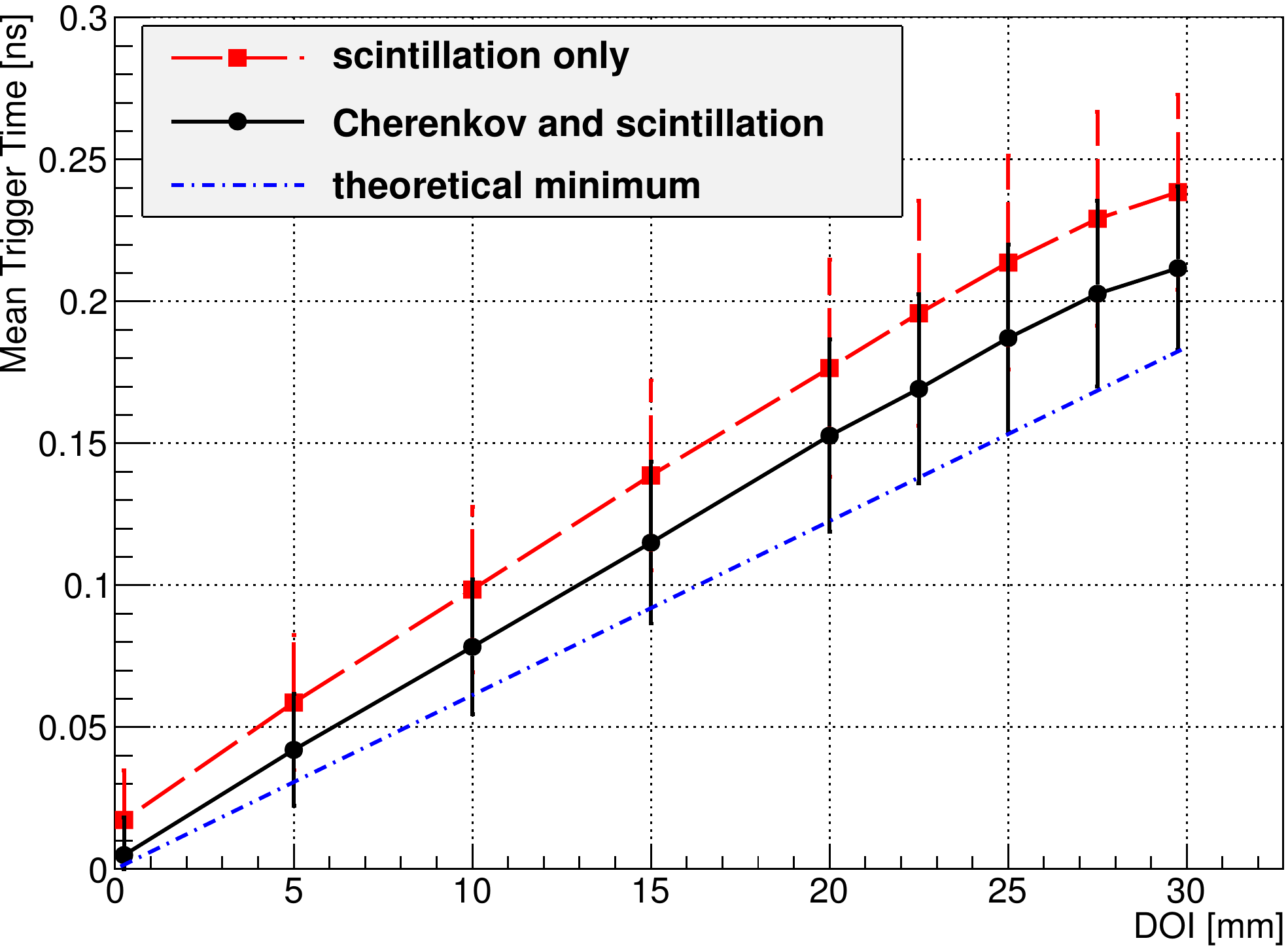}
\caption{Mean trigger time of the first photon vs. DOI for detection of only scintillation photons or detection of both. The blue dashed-dotted line indicates the theoretical minimum trigger time in LSO, when the photons undergo no reflections while propagating to the detector.}
\label{fig:MeanArrivalTime}
\end{figure}

\begin{figure}[htb] 
\centering 
\includegraphics[width=0.85\columnwidth,keepaspectratio]{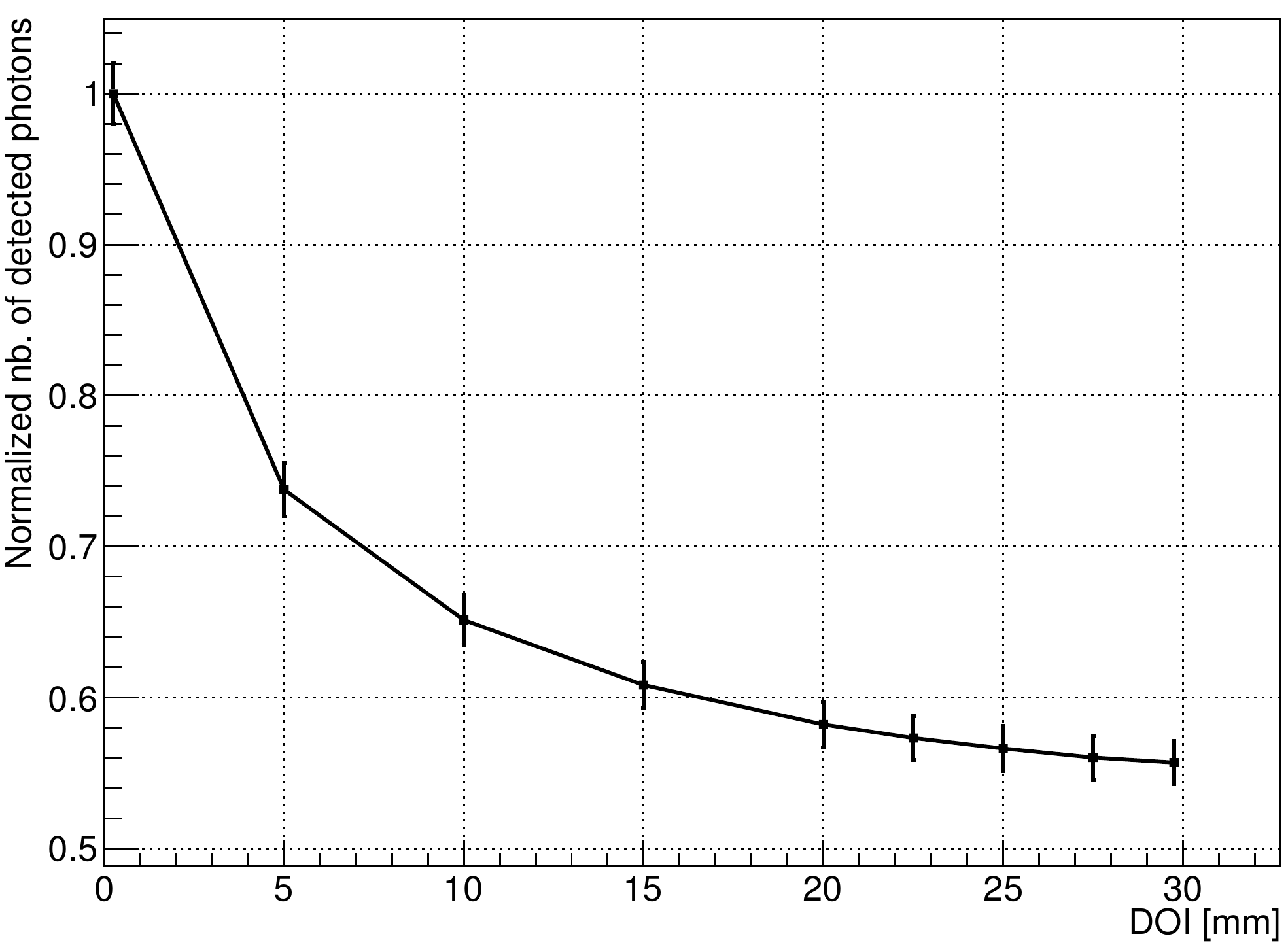}
\caption{Number of detected photons as a function of the DOI. This dependency can be used for time walk correction in PET-like detector configurations.}
\label{fig:NbOfPhotons}
\end{figure}

From the simulations, the mean trigger times at single photon level were determined for various DOIs, see figure \ref{fig:MeanArrivalTime}. Until large DOIs, the trigger times are depending linearly on the DOI, but deviate at DOIs $>$\,25\,mm. The reason is that for low DOIs, photons emitted towards the photon detector are triggering the detector, the photons emitted into the opposite direction arrive at the detector with a significant delay. With increasing DOI, the time delay of these two consecutive peaks decreases until they merge at high DOIs, resulting in a higher photon rate at the photon detector and, therefore, in earlier trigger probability \cite{Moses1999}. Furthermore, the number of detected photons is dependent on the DOI, see figure \ref{fig:NbOfPhotons}. It is clearly visible that the light output of the crystal decreases with increasing DOI.

Since Cherenkov photons provide very fast response to the photon interaction, for good time resolution it is beneficial to detect as many Cherenkov photons as possible. Unfortunately, many of them are lost in real detector systems due to low quantum efficiencies of photon detectors in the blue and UV-range and the cut-off frequencies of photon transmission in scintillators. Analyzing scintillation pulse shapes and detection of the first and second rise of photon arrivals, can provide information about DOI, and help reducing parallax errors of PET systems.

Due to limited time resolution of state-of-the-art photon detectors it is difficult to discriminate the first rise, the width of the plateau and the second rise of the photon arrival rate. Nevertheless, this effect should be observable as variation of the rise time of scintillation pulses. Therefore, measuring the rise time or the number of detected photons not only parallax errors can be reduced by estimation of the DOI but also improved time resolution by determination of a corrected time stamp of interactions of the 511\,keV photons inside the scintillator can be achieved. A dependency of scintillation rise times and the number of detected photons on the DOI has been measured by \cite{Wiener2010, Moses1999}.

\section{Time Resolution of a Scintillator}
\label{sec:TRLSO}
For investigations on the time resolution of a scintillator, a setup similar to that in figure \ref{fig:CoincidenceSetup} (b) was used, but with the only difference, that the photons were emitted from top, towards the photon detector. The time stamps were determined triggering on the first arriving photon (scintillation and Cherenkov effect). Contrary to the simulation in section \ref{sec:photontransport}, the DOI is unknown.

\begin{figure*}[htb] 
\centering 
\includegraphics[width=1.\textwidth,keepaspectratio]{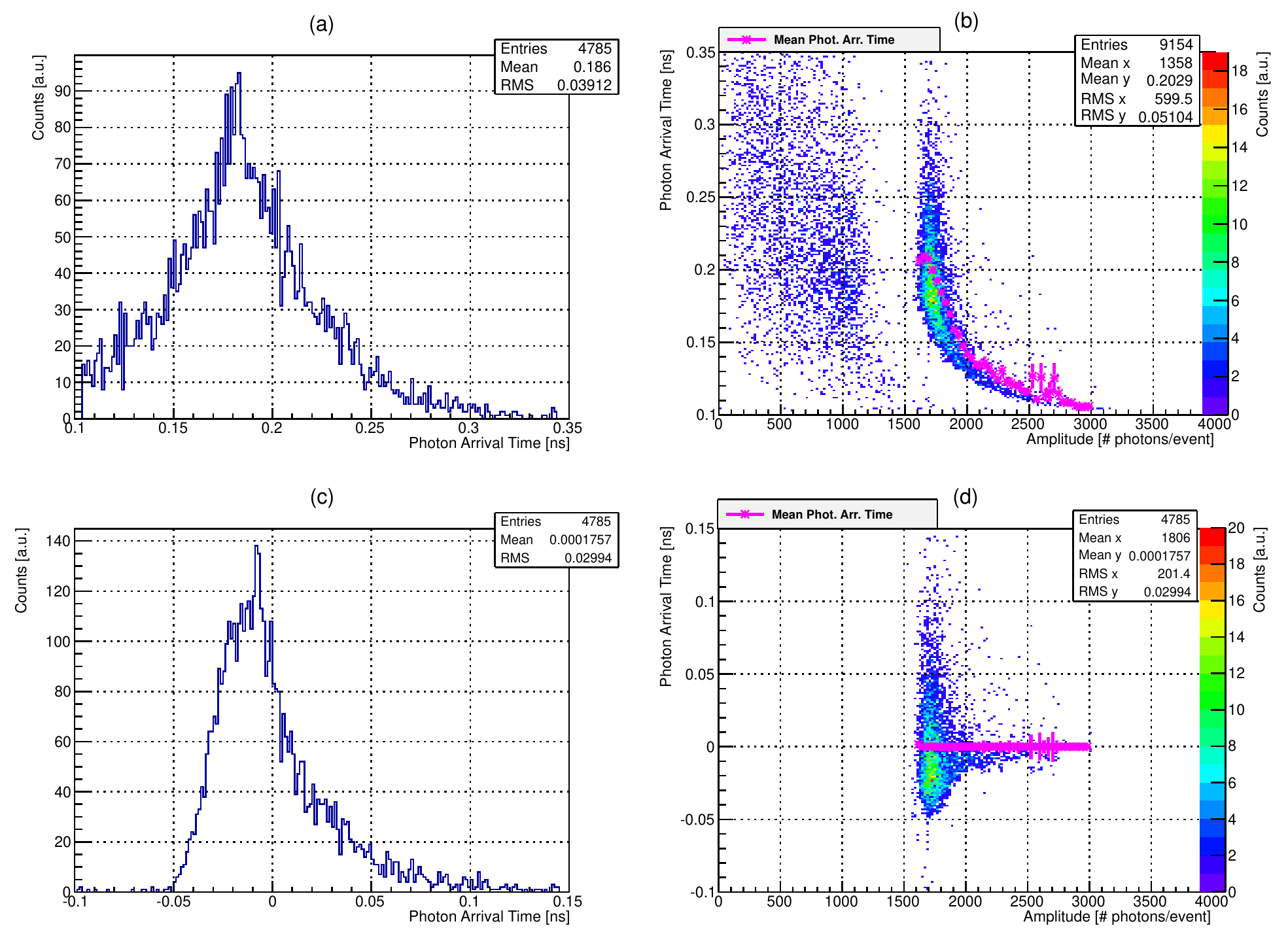}
\caption{Photon arrival times for the determination of the time resolution of a scintillator. (a) shows the photon arrival time, (b) the dependency of photon arrival time on number of arriving photons. In (b) also the Compton events are visible in the range below 1600 photons per event: the mean photon arrival times were calculated for different amplitudes and used to perform a time walk correction. Figures (c) and (d) show the time walk corrected photon arrival times and the corrected scatter plot, respectively.}
\label{fig:scatter}
\end{figure*}

The resulting photon arrival times can be seen in figure \ref{fig:scatter} (a). Figure \ref{fig:scatter} (b) shows a scatter plot of the number of optical photons per 511\,keV photon and the time when the first photon is arriving at the detector. Compton scattered events are discriminated with a discriminator threshold of 1600 photons per event. 92\,\% of the  511\,keV photons were detected, and 48\,\% of them were Compton events. In figure \ref{fig:scatter} (b) the time walk is visible. By calculating the mean photon arrival times for increasing amplitudes, time walk correction was applied. The corrected arrival time spectra can be seen in figure \ref{fig:scatter} (c) and (d).

The total time resolution $\sigma_{\textrm{\footnotesize{total}}}$ of the scintillator is $\cong$\,39\,ps. The corrected data of figure \ref{fig:scatter} (c) gives $\sigma_{\textrm{\footnotesize{corrected}}} \cong 30$\,ps. Using
\begin{equation*}
 \sigma_{\textrm{\footnotesize{total}}}^2 = \sigma_{\textrm{\footnotesize{time-walk}}}^2+\sigma_{\textrm{\footnotesize{corrected}}}^2
\end{equation*}
results in $\sigma_{\textrm{\footnotesize{time-walk}}}\cong25$\,ps.

The corrected time resolution $\sigma_{\textrm{\footnotesize{corrected}}}$ still includes the standard deviation of the scintillation process, the contribution of the Cherenkov process and a contribution of photon propagation. Note, that the name of the variables $\sigma$ do not imply normal distributions, but are measures for the standard variation. 

\section{Discussion and Conclusion}
Simulations for determination of the time resolution of LSO:Ce crystals were performed. The influence of crystal sizes on the CTR for PET-like detector systems was shown and ranged from 32\,ps to 144\,ps FWHM for crystal lengths of 1\,mm to 30\,mm. Including the detection of Cherenkov photons showed a significant improvement of the CTR (12 to 125\,ps FWHM). It was shown that the light output and the photon rate distribution at the photon detector are dependent on the DOI and that applying time walk correction significantly improves the time resolution from $\sigma_{\textrm{\footnotesize{total}}}\cong 39$\,ps to $\sigma_{\textrm{\footnotesize{corrected}}} \cong 30$\,ps. The contribution of the time walk was $\sigma_{\textrm{\footnotesize{time-walk}}}\cong25$\,ps.

By measuring the pulse amplitude or the rise time of a scintillation pulse, the DOI can be estimated. Information on the DOI allows to reduce parallax errors for PET and determination of the accurate time stamps of photon interactions results in improvement of TOF for PET. As the development of readout electronics proceeds quickly, extracting amplitude and rise-time information is realistic also for full TOF-PET systems \cite{Ashmanskas2011}.

For real PET-systems, photon detectors with high quantum efficiency in the blue- and UV-range and scintillators with increased transmission in these wavelength-bands would help to improve TOF for PET by making use of the Cherenkov effect. However, for time resolutions of state-of-the-art photon detectors the benefit from the Cherenkov effect is small, but becomes increasingly important if the time resolution of photon detectors approximates the time resolution of the scintillators. 

\section*{Acknowledegments}
This work was partly funded by EU-project HadronPhysics3 (project 283286).

%% The Appendices part is started with the command \appendix;
%% appendix sections are then done as normal sections
%% \appendix
%% \section{}
%% \label{}

\end{document}